\documentstyle[12pt]{article}
\begin{document}
\pagestyle{empty}
\setlength{\oddsidemargin}{0.5cm} \setlength{\evensidemargin}{0.5cm}
\setlength{\footskip}{1.5cm}
\renewcommand{\thepage}{-- \arabic{page} --}
% -------------------------------------------------------------------
\vspace*{-2.5cm}
\begin{flushright}
TOKUSHIMA 95-04 \\ (hep-ph/9510269) \\ October 1995
\end{flushright}
\vspace*{1.25cm}

\renewcommand{\thefootnote}{*)}
\centerline{\ {\large\bf EW Precision Analysis and the Higgs
Mass}\,\footnote{Invited Talk at the second German--Polish Symposium
``{\it New Ideas in the Theory of Fundamental Interactions}",
Zakopane, Poland, September 11-15, 1995 (to appear in the
Proceedings).}}

\vspace*{1.75cm}
\renewcommand{\thefootnote}{**)}
\centerline{\sc \phantom{**)}Zenr\=o HIOKI\,\footnote{E-mail:
hioki@ias.tokushima-u.ac.jp}}

\vspace*{1.75cm}
\centerline{\sl Institute of Theoretical Physics,\ University of
Tokushima}

\vskip 0.3cm
\centerline{\sl Tokushima 770,\ JAPAN}

\vspace*{3cm}
\centerline{ABSTRACT}

\vspace*{0.4cm}
\baselineskip=20pt plus 0.1pt minus 0.1pt
Two topics on the standard electroweak theory are discussed based on
its remarkable success in precision analyses. One is a test of
structure of the radiative corrections to the weak-boson masses as a
further precision analysis. The other is an indirect Higgs-boson
search through the radiative corrections to the various quantities
measured at LEP.
\vfill
\newpage
% -------------------------------------------------------------------
\renewcommand{\thefootnote}{\sharp\arabic{footnote}}
\renewcommand{\theequation}{\arabic{section}.\arabic{equation}}
% -------------------------------------------------------------------
\baselineskip=21.0pt plus 0.2pt minus 0.1pt
\pagestyle{plain} \setcounter{footnote}{0}
% 1111111111111111111111111111111111111111111111111111111111111111111
\section*{\normalsize\bf$\mbox{\boldmath $\S$}\!\!$
1. Introduction}
\setcounter{section}{1}\setcounter{equation}{0}

\vspace*{-0.3cm}
The discovery of the top quark \cite{TOP} has completed the fermion
world in the three generation scheme. In the framework of the
standard (minimal) electroweak theory, we now have only one
yet-undiscovered ingredient left: the Higgs boson. Combining this
with the fact that the electroweak theory (with the radiative
corrections) has been quite successful in precision analyses through
LEP, SLC, Tevatron and lots of other experimental information, we
find ourselves in a position to proceed to further more detailed
studies of this theory.

At this Symposium, I gave a talk about two topics under this
circumstance based upon some of our recent works \cite{ZH94,CH}: One
is a test of structure of the EW radiative corrections via $W/Z$
masses. The other is a Higgs-boson search through the radiative
corrections and precision LEP data. The latter has already been a
popular subject, and there are a lot of related papers (see [4 -- 6]
and references cited therein). I do not mean that we developed some
new technique to analyze the data. However, it is quite significant
for future experiments to draw any information on the Higgs mass, and
I showed some results which Consoli and I obtained lately.

Before stepping into actual discussions, let me briefly describe what
we have been studying on these topics. First, EW corrections consist
of several parts with different properties, and I examined via
$\alpha$, $G_F$ and $M_{W,Z}$ what would happen if each of them would
not exist. For example, there are top-quark corrections which do not
decouple, i.e., become larger and larger as $m_t$ increases. Studying
them are significant not only because it is a test of the EW theory
as a renormalizable field theory but also because the existence of
such effects is a characteristic feature of theories in which
particle masses are produced through spontaneous symmetry breakdown
plus large Yukawa couplings.

Next, on the Higgs search. Stimulated by the first CDF report on the
top-quark evidence, Najima and I considered if there is not any
problem in the EW theory. We then found that the Higgs mass needs to
be 1.1-1.2 TeV in order for $M_W|_{m_t=174\ {\rm GeV}}$ to reproduce
the central value of $M_W^{exp}$, contrary to some other analyses
using the LEP/SLC data which prefer a lighter Higgs boson: $m_{\phi}$
$\lower0.5ex\hbox{$\buildrel <\over\sim$}$ 300 GeV \cite{EFL,EFL95}.
At present, it is not that serious since such a lighter Higgs is also
allowed if we take into account the size of ${\mit\Delta}m_t^{exp}$
and ${\mit\Delta}M_W^{exp}$, but this motivated us to analyze the LEP
data our own way.

The first subject is discussed in section 2, and the second one is in
$\S\!\!$ 3. Section 4 is for brief summary and discussions.
% 2222222222222222222222222222222222222222222222222222222222222222222
\section*{\normalsize\bf$\mbox{\boldmath $\S$}\!\!$
2. Structure of EW Corrections}
\setcounter{section}{2}\setcounter{equation}{0}

\vspace*{-0.3cm}
Within the electroweak theory, the muon-decay width up to the
$O(\alpha)$ corrections is calculated as
\begin{eqnarray}% ---------------------------------------------------
{\mit\Gamma}
={\mit\Gamma}^{(0)}(\alpha, M_W, M_Z)\cdot(1+2{\mit\Delta}r),
\end{eqnarray}% -----------------------------------------------------
where ${\mit\Gamma}^{(0)}$ is the lowest-order width in terms of the
fine-structure constant $\alpha$ and the weak-boson masses $M_{W,Z}$,
and ${\mit\Delta}r$ is the corrections to the amplitude. As mentioned
in $\S\!\!$ 1, ${\mit\Delta}r$ consists of several parts with
different properties: the leading-log terms ${\mit\Delta}\alpha$, the
non-decoupling top-quark terms ${\mit\Delta}r[m_t]$ and the other
terms including the bosonic effects. On the other hand, its
experimental data, ${\mit\Gamma}^{exp}$, is usually expressed by the
Fermi coupling constant $G_F$. Therefore, by solving ${\mit\Gamma}=
{\mit\Gamma}^{exp}$ on $M_W$, we get
\begin{eqnarray}% ---------------------------------------------------
M_W=M_W(\alpha, G_F, M_Z, {\mit\Delta}r). \label{eq1}
\end{eqnarray}% -----------------------------------------------------
This formula, the $M_W$-$M_Z$ relation, is the main tool of my
analyses in this section.\footnote{Over the past several years, some
    corrections beyond the one-loop approximation have been computed.
    They are two-loop top-quark corrections \cite{BBCCV} and QCD
	corrections up to $O(\alpha_{\rm QCD}^2)$ for the top-quark loops
    \cite{HKl} (see \cite{FKS} as reviews). As a result, we have now
	a formula including $O(\alpha\alpha_{\rm QCD}^2 m_t^2)$ and
	$O(\alpha^2 m_t^4)$ effects. In	the following, $M_W$ is always
	computed by incorporating all of these higher-order terms as
	well, although I will express the whole corrections with these
	terms also as ${\mit\Delta}r$ for simplicity.}\ %----------------

Let me show first by using this formula how the theory with the full
corrections is successful, though it is already a well-known fact. We
thereby have
\begin{eqnarray}% ---------------------------------------------------
M_W^{(0)}=80.9404\pm 0.0027\ {\rm GeV\ \ and}\ \
M_W=80.36\pm 0.09\ {\rm GeV}
\end{eqnarray}% -----------------------------------------------------
for $M_Z^{exp}=91.1887\pm 0.0022$ GeV \cite{LEP}, where $M_W^{(0)}
\equiv M_W(\alpha, G_F, M_Z, {\mit\Delta}r=0)$ and $M_W$ is for
$m_t^{exp}=180 \pm 12$ GeV \cite{TOP}, $m_{\phi}=300$ GeV and
$\alpha_{\rm QCD}(M_Z)$=0.118. Concerning the uncertainty of $M_W$,
0.09 GeV, I have a little overestimated for safety. From these
results, we can find that the theory with the corrections is in good
agreement with the experimental value $M_W^{exp}=80.26\pm 0.16$ GeV
\cite{wmass}, while the tree prediction fails to describe it at about
4.3$\sigma$ (99.998 \% C.L.).

We are now ready. First, let us see if taking only ${\mit\Delta}
\alpha(\sim\alpha\ln(m_f/M_Z))$ into account is still a good
approximation (``(QED-)improved-Born approximation"), which was shown
to be quite successful in \cite{NOV93}. The $W$ mass is calculated
within this approximation by putting ${\mit\Delta}r=0$ and replacing
$\alpha$ with $\alpha(M_Z)(=\alpha/(1-{\mit\Delta}\alpha))$ in
Eq.(\ref{eq1}), where $\alpha(M_Z)=1/(128.92\pm 0.12)$.\footnote{
   Recently three papers appeared in which $\alpha(M_Z)$ is
   re-evaluated from the data of the total cross section of $e^+ e^-
   \rightarrow\gamma^*\rightarrow hadrons$ \cite{alphamz} (their
   updated results are given in \cite{tatsu}). Here I simply took the
   average of the maximum and minimum among them.}\ %----------------
The result is
\begin{eqnarray}% ---------------------------------------------------
M_W[{\rm Born}] =79.964\pm 0.017~{\rm GeV},
\end{eqnarray}% -----------------------------------------------------
which leads to
\begin{eqnarray}% ---------------------------------------------------
M_W^{exp}-M_W[{\rm Born}]~=~0.30\pm 0.16~{\rm GeV}.
\end{eqnarray}% -----------------------------------------------------
This means that $M_W[{\rm Born}]$ is in disagreement with the data
now at $1.9\sigma$, which corresponds to about 94.3 \%\ C.L..
Although the precision is not yet sufficiently high, it indicates
some non-Born terms are needed which give a positive contribution to
the $W$ mass. It is noteworthy since the electroweak theory predicts
such positive non-Born type corrections unless the Higgs is extremely
heavy (beyond TeV scale). Similar analyses were made also in
\cite{NOV94}.

The next test is on the non-decoupling top-quark effects. Except for
the coefficients, their contribution to ${\mit\Delta}r$ is
\begin{eqnarray}% ---------------------------------------------------
{\mit\Delta}r[m_t]\sim \alpha (m_t/M_Z)^2+\alpha\ln(m_t/M_Z).
\end{eqnarray}% -----------------------------------------------------
According to my strategy, I computed the $W$ mass by using the
following ${\mit\Delta}r'$ instead of ${\mit\Delta}r$ in
Eq.(\ref{eq1}):
\begin{eqnarray}% ---------------------------------------------------
{\mit\Delta}r'\equiv {\mit\Delta}r-{\mit\Delta}r[m_t].
\end{eqnarray}% -----------------------------------------------------
The resultant $W$ mass is denoted as $M_W'$. The important point is
to subtract not only $m_t^2$ term but also $\ln(m_t/M_Z)$ term,
though the latter produces only very small effects as long as $m_t$
is not extremely large. ${\mit\Delta}r'$ still includes $m_t$
dependent terms, but no longer diverges for $m_t\to +\infty$ thanks
to this subtraction. I found that $M_W'$ takes the maximum value for
the largest $m_t$ and the smallest $m_{\phi}$. That is, we get an
inequality
\begin{eqnarray}% ---------------------------------------------------
M_W'\ \leq\ M_W'[m_t^{max}, m_{\phi}^{min}].
\end{eqnarray}% -----------------------------------------------------

We can use $m_t^{exp}=180\pm 12$ GeV \cite{TOP} and $m_{\phi}^{exp}>
65.1$ GeV \cite{Higgs} in the right-hand side of the above
inequality, i.e., $m_t^{max}=180+12$ GeV and $m_{\phi}^{min}=65.1$
GeV, but I first take $m_t^{max}\to +\infty$ and $m_{\phi}^{min}=0$
in order to make the result as data-independent as possible. The
accompanying uncertainty for $M_W'$ is estimated at most to be about
0.03 GeV. We have then
\begin{eqnarray}% ---------------------------------------------------
M_W' < 79.950 (\pm 0.030) \ {\rm GeV\ \ and\ \ }
M_W^{exp}-M_W' > 0.31\pm 0.16\ {\rm GeV},
\end{eqnarray}% -----------------------------------------------------
which show that $M_W'$ is in disagreement with $M_W^{exp}$ at about
$1.9\sigma$. This means that 1) the electroweak theory is not able to
be consistent with $M_W^{exp}$ {\it whatever values $m_t$ and
$m_{\phi}$ take} if ${\mit\Delta}r[m_t]$ would not exist, and 2) the
theory with ${\mit\Delta}r[m_t]$ works well, as shown before, for
experimentally-allowed $m_t$ and $m_{\phi}$. Combining them, we can
summarize that the latest experimental data of $M_{W,Z}$ demand, {\it
independent of $m_{\phi}$}, the existence of the non-decoupling
top-quark corrections. The confidence level of this result becomes
higher if we use $m_t^{max}=180+12$ GeV and $m_{\phi}^{min}=65.1$
GeV:
\begin{eqnarray}% ---------------------------------------------------
M_W' < 79.863 (\pm 0.030) \ {\rm GeV\ \ and\ \ }
M_W^{exp}-M_W' > 0.40\pm 0.16\ {\rm GeV},
\end{eqnarray}% -----------------------------------------------------
that is, $2.5\sigma$ level.

Finally, let us look into the bosonic contribution. It was pointed
out in \cite{DSKK} by using various high-energy data that such
bosonic electroweak corrections are now required. I studied whether
we could observe a similar evidence in the $M_W$-$M_Z$ relation. For
this purpose, we have to compute $M_W$ taking account of only the
pure-fermionic corrections ${\mit\Delta}r[f]$. Since ${\mit\Delta}
r[f]$ depends on $m_t$ strongly, it is not easy to develop a
quantitative analysis of it without knowing $m_t$. Therefore, I took
into account $m_t^{exp}$ from the beginning in this case. I express
thus-computed $W$ mass as $M_W[{\rm f}]$. The result became
\begin{eqnarray}% ---------------------------------------------------
M_W[{\rm f}]=80.48\pm 0.09\ {\rm GeV}.
\end{eqnarray}% -----------------------------------------------------
This value is of course independent of the Higgs mass, and leads to
\begin{eqnarray}% ---------------------------------------------------
M_W[{\rm f}]-M_W^{exp}=0.22\pm 0.18\ {\rm GeV}, \label{eq2}
\end{eqnarray}% -----------------------------------------------------
which tells us that some non-fermionic contribution is necessary at
$1.2\sigma$ level.

It is of course too early to say from Eq.(\ref{eq2}) that the bosonic
effects were confirmed in the $M_W$-$M_Z$ relation. Nevertheless,
this is an interesting result since we could observe nothing before:
Actually, the best information on $m_t$ before the first CDF report
(1994) was the bound $m_t^{exp}>$ 131 GeV by D0 \cite{D0}, but we can
thereby get only $M_W[{\rm f}]>$ 80.19 ($\pm$0.03) GeV while
$M_W^{exp}[94]$ was 80.23$\pm$0.18 GeV (i.e., $M_W[{\rm f}]-M_W^{exp}
>-0.04\pm0.18$ GeV). We will be allowed therefore to conclude that
``the bosonic effects are starting to appear in the $M_W$-$M_Z$
relation thanks to the discovery of the top-quark".
% 3333333333333333333333333333333333333333333333333333333333333333333
\section*{\normalsize\bf$\mbox{\boldmath $\S$}\!\!$
3. Indirect Higgs Search}
\setcounter{section}{3}\setcounter{equation}{0}

\vspace*{-0.3cm}
Here I wish to discuss what information on the Higgs we can get from
precision LEP data. As a matter of fact, it is not that easy to draw
its indirect information from existing experimental data since the
Higgs mass $m_{\phi}$ enters EW radiative corrections only
logarithmically at one-loop level \cite{Vel}. Therefore, at present,
one can only hope to separate out the heavy Higgs-mass range (say
$m_{\phi}\sim$ 500-1000 GeV) from the low mass regime $m_{\phi}
\sim$100 GeV as predicted, for instance, from supersymmetric
theories. Such analyses are, however, still very important and
indispensable for future experiments at, e.g., LHC/NLC.

For our analysis, we used in \cite{CH} the disaggregated data, just
as presented by the experimental Collaborations, without taking any
average of the various results. This type of analysis is interesting
by itself to point out the indications of the various sets of data
since even a single measurement, if sufficiently precise, can provide
precious information. At the same time, since the LEP data are
becoming so precise, before attempting any averaging procedure one
should first analyze the various measurements with their errors and
check that the distribution of the results fulfills the requirements
of Gaussian statistics. Without this preliminary analysis, one may
include uncontrolled systematic effects which can sizably affect the
global averages.

We first restricted to a fixed value of the top-quark mass $m_t=180$
GeV. As input data, we used the available, individual results
${\mit\Gamma}_Z$, $\sigma_{had}$, $R_{\ell}$, $A_{\rm FB}(\ell)$ and
$A_{e,\tau}$ from the four Collaborations as quoted in \cite{LEP},
where $R_{\ell}\equiv{\mit\Gamma}_{had}/{\mit\Gamma}_{\ell}$ and
$A_{\ell}\equiv 2g_V^{\ell}g_A^{\ell}/\{(g_V^{\ell})^2+(g_A^{\ell})^2
\}$ ($\ell=e,\mu,\tau$, and $g_{V,A}^{\ell}$ are the vector and
axial-vector couplings of $\ell$ to $Z$).\footnote{We did not
   consider the LR asymmetry by SLD \cite{SLD} since it is already
   known that it demands a very heavy top (around 240 GeV) when the
   lower bound on $m_{\phi}$ is taken into account, or conversely
   $m_{\phi}$ must be much lower than this bound when $m_t^{exp}$ is
   used within the standard EW theory (see, e.g., Ellis et al. in
   \cite{EFL}).}% ---------------------------------------------------
The theoretical computations have been performed with the computer
code TOPAZ0 \cite{TOPAZ0}, and the main results are given in Table 1.
There we do not see any specific indication on $m_{\phi}$, but a
heavy Higgs seems to be a little bit favored by the total $\chi^2$.

\vskip -0.15cm
\centerline{\bf ------------------------------}
\centerline{\bf Table 1}
\centerline{\bf ------------------------------}

We, however, found some problems in the $\tau$ forward-backward
asymmetry as shown below. Let us consider the global averages
\begin{eqnarray}% ---------------------------------------------------
&&A^{exp}_{\rm FB}(e)=0.0154 \pm 0.0030,      \label{eq3}\\
&&A^{exp}_{\rm FB}(\mu)=0.0160 \pm 0.0017,    \label{eq4}\\
&&A^{exp}_{\rm FB}(\tau)=0.0209 \pm 0.0024,   \label{eq5}
\end{eqnarray}% -----------------------------------------------------
and transform the averages for $A_e$ and $A_{\tau}$
\begin{eqnarray}% ---------------------------------------------------
A^{exp}_e=0.137\pm 0.009,~~~A^{exp}_{\tau}=0.140\pm 0.008
\end{eqnarray}% -----------------------------------------------------
into ``effective" F-B asymmetries by using
\begin{eqnarray}% ---------------------------------------------------
A_{\rm FB}(e)={3\over 4}({A_e})^2,~~~
A_{\rm FB}(\tau)={3\over 4}A_e A_{\tau}, \label{eq6}
\end{eqnarray}% -----------------------------------------------------
which hold in the electroweak theory. We find
\begin{eqnarray}% ---------------------------------------------------
A^{eff}_{\rm FB}(e)=0.0141\pm0.0019,~~~
A^{eff}_{\rm FB}(\tau)=0.0144\pm0.0018
\label{eq7}
\end{eqnarray}% -----------------------------------------------------
in very good agreement with Eqs.(\ref{eq3}) and (\ref{eq4}) but not
with Eq.(\ref{eq5}). Therefore, there might be some problem in the
direct measurement of $A_{\rm FB}(\tau)$ since all other measurements
are in excellent agreement with each other.

Just to have an idea of the effect, we computed the $\chi^2$ without
$A_{\rm FB}^{exp}(\tau)$. The results are illustrated in Table 2,
which should be compared with Table 1. We find that the tendency
toward a heavy Higgs becomes stronger and the best values of the
$\chi^2$ are obtained for a large value of $m_{\phi}$, just as in the
case of the $W$ mass mentioned in $\S\!\!$ 1. It is still not easy to
get a definite conclusion from this, but the ``bulk" of the LEP data,
namely those well consistent with each other, show no preference for
a light Higgs boson, to say the least of it.

\vskip -0.15cm
\centerline{\bf ------------------------------}
\centerline{\bf Table 2}
\centerline{\bf ------------------------------}

Finally, to see the $m_t$-dependence of $\chi^2$, I show in Tables
3 and 4 the total $\chi^2$ for $m_t$=170, 180 and 190 GeV including
all data or excluding $A^{exp}_{\rm FB}(\tau)$. By increasing
(decreasing) the top-quark mass, a larger (smaller) value of
$m_{\phi}$ comes to be favored. This is because the leading top and
Higgs terms in the radiative corrections have opposite signs to each
other. For a heavier top
$m_t\ \lower0.5ex\hbox{$\buildrel >\over\sim$}$
180 GeV, however, Tables 3 and 4 give rather different information
and it becomes crucial to include the problematic data for
$A^{exp}_{\rm FB}(\tau)$ to accommodate $m_{\phi}\sim$ 100 GeV.

\vskip -0.15cm
\centerline{\bf ------------------------------}
\centerline{\bf Tables 3 and 4}
\centerline{\bf ------------------------------}

We have no mind to claim that Tables 2 and 4 represent a more
faithful representation of the real physical situation than Tables 1
and 3. Most likely, our results suggest only that further improvement
in the data taking is necessary for a definitive answer. We may,
however, conclude thereby that it is not a good idea to focus on a
light-mass region in Higgs searches at future experiments.
% 4444444444444444444444444444444444444444444444444444444444444444444
\section*{\normalsize\bf$\mbox{\boldmath $\S$}\!\!$
4. Summary and Discussions}
\setcounter{section}{4}\setcounter{equation}{0}

\vspace*{-0.3cm}
Let us briefly summarize and discuss what I have talked. In section
2, I have shown that we can now test not only (1) the whole EW
corrections but also their various parts separately: (2) the
light-fermion leading-log corrections which lead to the improved-Born
approximation, (3) the non-decoupling $m_t$ corrections and (4) the
bosonic corrections. Studying corrections (1) is a test of the theory
as a renormalizable field theory, while (2)$\sim$(4) are more
detailed tests.

The improved-Born approximation succeeded to a certain extent, which
is related to the fact that the EW theory unifies the weak
interaction (with $q^2\simeq M_W^2$ scale) and the electromagnetic
interactions (with $q^2\simeq 0$ scale). We, however, have seen that
some non-Born corrections are now starting to appear. Next we
observed that the non-decoupling $m_t$ corrections are also required,
which gives a strong support to the mechanism that $m_t$ is produced
via spontaneous symmetry breakdown plus large Yukawa couplings.
Similar way we also tested the bosonic corrections and found some
small indication for them.

Based on this excellent success, we are able to explore the remaining
unknown area, i.e., the Higgs sector, which I discussed in section 3.
Concerning such a Higgs search, there are already a lot of papers.
Unfortunately, the $m_{\phi}$-dependence of one-loop quantities are
only logarithmic in the minimal scheme and therefore it is not easy
to get any strong restriction on $m_{\phi}$, but we have so far
obtained some quantitative information. Such information is of course
extremely important for future experimental projects like LHC/NLC.
Several papers pointed out that the Higgs will be rather light, say
less than about 300 GeV \cite{EFL}. We have found, however, there is
also an indication for a rather heavy Higgs through our analysis of
LEP and $W$-mass data.

Due to the reason mentioned above, our results cannot be strong
either, but at least we can say it is risky to concentrate our
attention on a light-mass region in Higgs searches at future
experiments, though I am not a fan of a heavy Higgs boson. More
precise measurements of the top-quark and $W$-boson masses are
considerably significant for studying this problem (and also for
searching any new-physics effects), and I wish to expect that the
Tevatron and LEP II will give us a good answer for it in the very
near future.

\vskip 20pt
\vspace*{0.3cm}
\centerline{ACKNOWLEDGEMENTS}

\vspace*{0.3cm}
I am grateful to
R. R${{{\rm a}_{}}_{}}_{\hskip -0.18cm\varsigma}$czka
and the organizing committee of the Symposium for the invitation
and their warm hospitality. I also would like to thank M. Consoli and
R. Najima for collaboration, on which many parts of this talk is
based. During the Symposium, I enjoyed valuable conversation with
R. R${{{\rm a}_{}}_{}}_{\hskip -0.18cm\varsigma}$czka, G. Weiglein
and many other participants, which I appreciate very much. I must not
forget to thank Jan Fischer, without whom I could not have reached
the Symposium site on the first day. This work is supported in part
by the Grant-in-Aid for Scientific Research (No. 06640401) from the
Ministry of Education, Science, Sports and Culture, Japan.

\vskip 0.8cm
% RRRRRRRRRRRRRRRRRRRRRRRRRRRRRRRRRRRRRRRRRRRRRRRRRRRRRRRRRRRRRRRRRRR

% TTTTTTTTTTTTTTTTTTTTTTTTTTTTTTTTTTTTTTTTTTTTTTTTTTTTTTTTTTTTTTTTTTT
\newpage
\renewcommand{\arraystretch}{1.4}

\vspace*{0.7cm}
\begin{center}
\begin{tabular}{lcccc}
\ $\alpha_{\rm QCD}(M_Z)$
&~~~~ {$0.113$} &~~~~ {$0.125$} &~~~~ {$0.127$} & \ ~~ {$0.130$} \\
\ $m_{\phi}$(GeV)
&~~~~ {$100$}   &~~~~ {$100$}   &~~~~ {$500$}   & \ ~~ {$1000$}  \\
\hline\hline
ALEPH  &~~~~$11.2$ & ~~~~$15.2$ &~~~~$13.9$ &~~~~$14.7$ \\ \hline
DELPHI &~~~~$ 5.1$ & ~~~~$ 7.9$ &~~~~$ 6.7$ &~~~~$ 7.2$ \\ \hline
L3     &~~~~$11.6$ & ~~~~$ 6.0$ &~~~~$ 8.0$ &~~~~$ 9.2$ \\ \hline
OPAL   &~~~~$19.4$ & ~~~~$13.9$ &~~~~$ 8.5$ &~~~~$ 6.9$ \\ \hline
\hline
Total $\chi^2$
       &~~~~$47.3$ & ~~~~$43.0$ &~~~~$37.1$ &~~~~$38.1$ \\ \hline
\end{tabular}
\end{center}

\vskip 1cm \noindent
{\bf Table 1.}\ Total $\chi^2$ for the four Collaborations.
%--------------------------------

\vspace{2.5cm}
\begin{center}
\begin{tabular}{lcccc}
\ $\alpha_{\rm QCD}(M_Z)$
&~~~~ {$0.113$} &~~~~ {$0.125$} &~~~~ {$0.127$} & \ ~~ {$0.130$} \\
\ $m_{\phi}$(GeV)
&~~~~ {$100$}   &~~~~ {$100$}   &~~~~ {$500$}   & \ ~~ {$1000$}  \\
\hline\hline
ALEPH  &~~~~$10.2$ & ~~~~$14.3$ &~~~~$12.1$ &~~~~$12.5$ \\ \hline
DELPHI &~~~~$ 4.7$ & ~~~~$ 7.5$ &~~~~$ 5.9$ &~~~~$ 6.2$ \\ \hline
L3     &~~~~$ 8.4$ & ~~~~$ 2.8$ &~~~~$ 3.9$ &~~~~$ 4.8$ \\ \hline
OPAL   &~~~~$19.4$ & ~~~~$13.8$ &~~~~$ 8.0$ &~~~~$ 6.1$ \\ \hline
\hline
Total $\chi^2$
       &~~~~$42.7$ & ~~~~$38.4$ &~~~~$29.9$ &~~~~$29.6$ \\ \hline
\end{tabular}
\end{center}

\vskip 1cm \noindent
{\bf Table 2.}\ Total $\chi^2$ for the four Collaborations by
excluding the data for $A^{exp}_{\rm FB}(\tau)$.
%--------------------------------------------------------------------
\newpage
\vspace*{-1cm}
\begin{center}
ALEPH+DELPHI+L3+OPAL

\vskip 30 pt
\begin{tabular}{ccccc}
$\alpha_{\rm QCD}(M_Z)$
&~~~~ {$0.113$} &~~~~ {$0.125$} & \ ~~ {$0.127$} &~~~~ {$0.130$} \\
$m_{\phi}$(GeV) &~~~~ {$100$} &~~~~ {$100$} &~~~~ {$500$} &
\ ~~ {$1000$} \\ \hline\hline
$m_t$(GeV)=
170~~~ &~~~~$46.3$ & ~~~~$38.4$ &~~~~$38.3$ &~~~~$41.2$ \\ \hline
\phantom{$m_t$(GeV)}=
180~~~ &~~~~$47.3$ & ~~~~$43.0$ &~~~~$37.1$ &~~~~$38.1$ \\ \hline
\phantom{$m_t$(GeV)}=
190~~~ &~~~~$51.8$ & ~~~~$50.4$ &~~~~$38.9$ &~~~~$37.5$ \\ \hline
\end{tabular}
\end{center}

\vskip 1cm \noindent
{\bf Table 3.}\ Total $\chi^2$ for the four Collaborations at various
values of $m_t$.
%--------------------------------

\vspace{2cm}
\begin{center}
ALEPH+DELPHI+L3+OPAL

\vskip 30 pt
\begin{tabular}{ccccc}
$\alpha_{\rm QCD}(M_Z)$
&~~~~ {$0.113$} &~~~~ {$0.125$} & \ ~~ {$0.127$} &~~~~ {$0.130$} \\
$m_{\phi}$(GeV) &~~~~ {$100$} &~~~~ {$100$} &~~~~ {$500$} &
\ ~~ {$1000$} \\ \hline\hline
$m_t$(GeV)=
170~~~ &~~~~$40.7$ & ~~~~$32.8$ &~~~~$29.7$ &~~~~$31.2$ \\ \hline
\phantom{$m_t$(GeV)}=
180~~~ &~~~~$42.7$ & ~~~~$38.4$ &~~~~$29.9$ &~~~~$29.6$ \\ \hline
\phantom{$m_t$(GeV)}=
190~~~ &~~~~$50.1$ & ~~~~$46.6$ &~~~~$32.9$ &~~~~$30.3$ \\ \hline
\end{tabular}
\end{center}

\vskip 1cm \noindent
{\bf Table 4.}\ Total $\chi^2$ for the four Collaborations at various
values of $m_t$ by excluding the data for $A^{exp}_{\rm FB}(\tau)$.
%--------------------------------------------------------------------
\end{document}